\documentclass[aps,prl,twocolumn,showpacs,preprintnumbers,amsmath,amssymb,
floatfix,superscriptaddress]{revtex4-2}

\usepackage{graphicx}
\usepackage{dcolumn}
\usepackage{bm}
\usepackage[usenames,dvipsnames]{color} 

\newcommand{\ybz}{\ensuremath{^{172}\text{Yb}^+ \,}}
\newcommand{\ybd}{\ensuremath{^{173}\text{Yb}^+ }}
\newcommand{\yb}{\ensuremath{\text{Yb}^+ \,}}

\newcommand{\affA}{Physikalisch-Technische Bundesanstalt, Bundesallee 100, 38116
Braunschweig, Germany}
\newcommand{\affB}{Institut f\"ur Quantenoptik, Leibniz
Universit\"at Hannover, Welfengarten 1, 30167 Hannover, Germany}

\begin{document}

\preprint{APS/123-QED}

\title{Precision spectroscopy in \yb ions}

\author{Jialiang Yu}\affiliation{\affA}
  \email{jialiang.yu@ptb.de}
\author{Kai C. Grensemann}\affiliation{\affA}\affiliation{\affB}
\author{Chih-Han Yeh}\affiliation{\affA}
\author{Ikbal Ahamed Biswas}\affiliation{\affA}
\author{Abhilasha Singh}\affiliation{\affA}
\author{Laura S. Dreissen}\affiliation{\affA}
\author{Henning A. Fr\"urst}\affiliation{\affA}

\author{Tanja E. Mehlst\"aubler}\affiliation{\affA}\affiliation{\affB}
 \email{tanja.mehlstaeubler@ptb.de}



\date{\today}

\begin{abstract}
we report on our progress on multi-ion spectroscopy with \yb ions. We first characterized the spatial homogeneity of the magnetic- and the rf-field. Subsequently, to equalize the immense AC Stark shifts of the individual ions in a Coulomb crystal, we utilize a holographic phase plate to convert the Gaussian beam into a flat-top shape. These efforts enable us to excite an 8-ion crystal (\ybz) on the highly forbidden electric octupole transition. Using \ybd with reduced AC Stark shift, the same beam profile could support the multi-ion operation of 20 ions. This paves the way for future multi-ion optical clocks
\end{abstract}

\maketitle



\section{Introduction}
Precision spectroscopy using atoms or molecules has emerged as an exceptionally sensitive tool for tests of physics beyond the Standard Model. Due to their extraordinary high precision, these laboratory-scale experiments can offer competitive constraints on new physics compared with those performed with high-energy colliders and cosmological observations. Among the various atomic species, Yb$^+$ has several advantages, e.g. electric octupole (E3) transition with nHz linewidth \cite{langeLifetime722021}, high sensitivity to local Lorentz violation (LLI) \cite{dzubaStronglyEnhancedEffects2016, shanivNewMethodsTesting2018a} and to the fluctuation of the fine structure constant \cite{flambaumSearchVariationFundamental2009}. We use \yb for testing of local Lorentz invariance in the electron-photon sector \cite{dreissenImprovedBoundsLorentz2022}, and to put limits on the 5th proposed forces between neutrons and electrons \cite{doorProbingNewBosons2025a}. These two experiments represent the highest accuracy to date in atomic systems and could be further improved by multi-ion operation. Increasing the number of ions has also been proven to be a necessary and successful way to improve the sensitivity of clocks and reduce the long averaging times \cite{herschbachLinearPaulTrap2012b, kellerControllingSystematicFrequency2019a, steinelEvaluationSr882023,pelzerMultiIonFrequencyReference2024, hausser115In+172YbCoulombCrystal2024, leibrandtProspectsThousandIonSn22024}, which has recently been successfully demonstrated in In$^+$ and Sr$^+$  ions \cite{ kellerControllingSystematicFrequency2019a, steinelEvaluationSr882023,pelzerMultiIonFrequencyReference2024, hausser115In+172YbCoulombCrystal2024}.  A significant obstacle to multi-ion operation is the strong AC Stark shift of the E3 transition, which is expected to be overcome with a highly promising candidate – \ybd.

\section{Towards multi-ion clock and LLI test}
To perform the test of Lorentz violation or operate optical clocks with multiple ions, simultaneous excitation of all ions on the highly forbidden E3 transition (from $^2S_{1/2}$ to $^2F_{7/2}$) is required. For \yb, the E3 transition frequency could be shifted unequally for the individual ion for several reasons, e.g., magnetic field gradient via Zeeman effect, and light intensity gradient via AC Stark shift. For the test of LLI, the gradient in the radio-frequency (RF) field that couples the Zeeman states of the $^2F_{7/2}$ manifold could degrade the fidelity of dynamical decoupling \cite{yehRobustScalableRf2023}. Here, we characterize these effects and suppress them to enable the multi-ion operation.

\subsection{Magnetic field homogeneity}

The resolved linewidth of the E3 transition is typically below 10 Hz. For even isotopes of \yb ions, this linewidth corresponds to a magnetic field fluctuation of about 20 µG, as they don’t have magnetic insensitive transitions. In odd isotopes, due to nuclear spin, first-order Zeeman insensitive transitions are available. However, for the LLI test, a RF pulse sequence is used to distribute the electron population across the Zeeman manifolds \cite{dreissenImprovedBoundsLorentz2022}. In this case, magnetic field gradients would lead to an unequal energy spacing between Zeeman substates for different ions within a crystal, i.e., frequency detuning of the RF pulse and therefore degrade the fidelity of the dynamical decoupling sequence. 

\begin{figure}[h]
	\centering\includegraphics[width=0.45\textwidth]{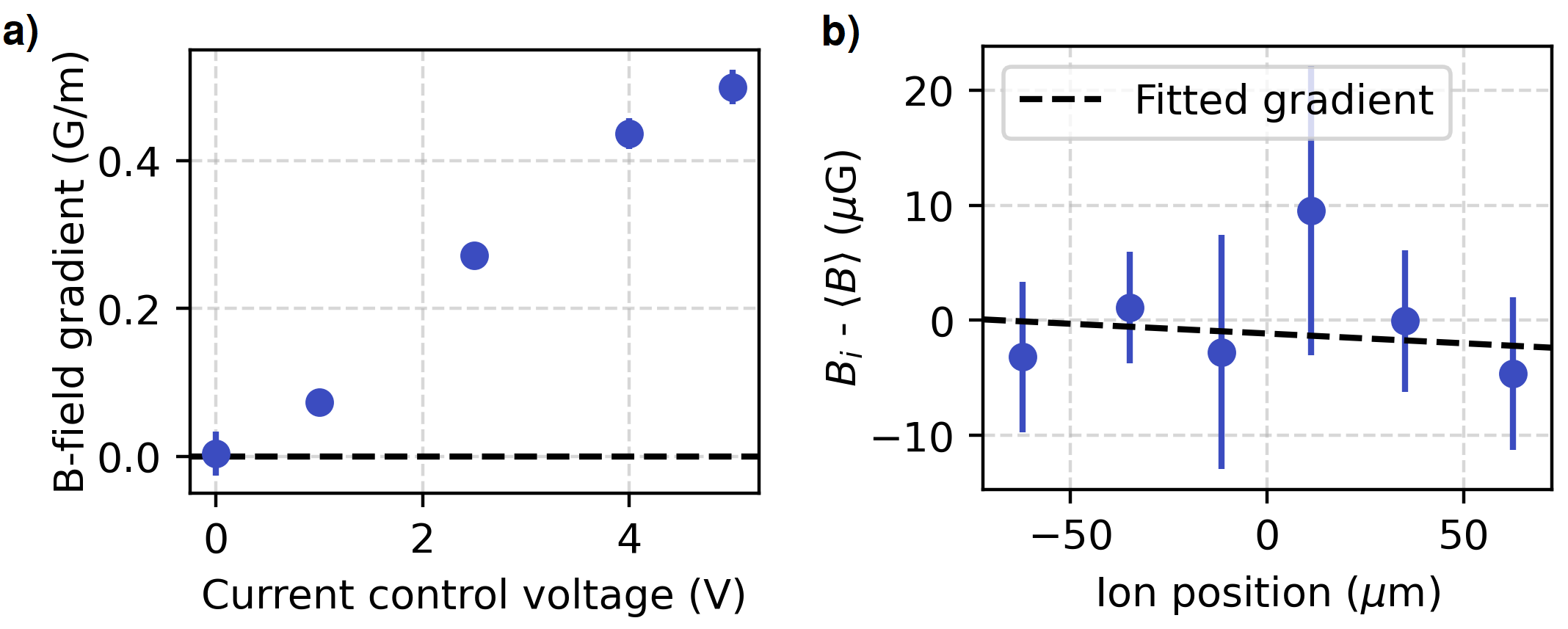}
	\caption{\label{fig1}
        	a) Magnetic field gradient at different control voltages. b) The achieved magnetic field gradient across an 80 $\mathrm{\mu m}$ ion crystal
        }
\end{figure}

In our setup, three pairs of copper coils are used to compensate for the stray magnetic field and generate the B-field along the quantization axis. One of the coils along the quantization axis is equipped with a control unit, which can fine adjust the electric current in this coil. By changing the control unit voltage, the magnetic field gradient at the ion can be fine minimized (Fig. \ref{fig1}a). The absolute magnetic field is stabilized via three additional coil pairs, fed with the signal from a flux gate sensor close to the ion trap.  We use the electric quadrupole transition at 411 nm as a quantum sensor to measure the magnetic field in the crystal. The achieved minimal magnetic gradient is shown in Fig. \ref{fig1}b. For the even \yb isotopes, this gradient corresponds to an RF frequency offset of 10 Hz over the whole crystal on the E3 transition. For the LLI test, this gradient would lead to a frequency detuning of 0.04

\subsection{Rf field homogeneity}
\begin{figure}[h]
	\centering\includegraphics[width=0.45\textwidth]{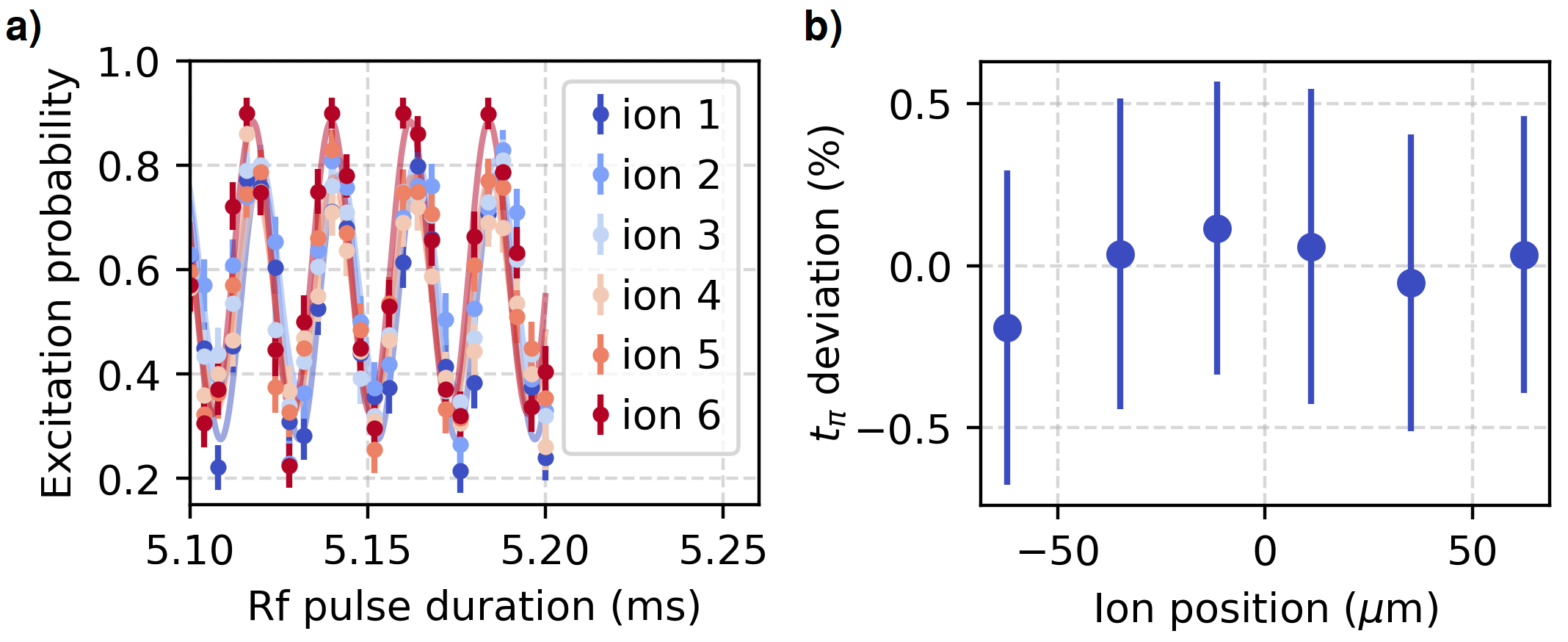}
	\caption{\label{fig2}
a) Rabi flop on the ground state Zeeman sublevels. b) Extrapolated pulse duration error for the dynamical decoupling sequence.
        }
\end{figure}

As described in the previous section, RF pulses are essential for the LLI test. While a magnetic field gradient causes to frequency detuning of the dynamical decoupling sequence, an RF field gradient leads to pulse time error. To characterize the RF field homogeneity, we drive the transition between the Zeeman manifold of the $^2S_{1/2}$ ground state (Fig. \ref{fig2}a). From the difference in Rabi frequency is a direct measure of pulse time error as shown in Fig. \ref{fig2}b. The error of ±0.25\% can be easily compensated by the UR10 sequence.

\subsection{Light intensity homogeneity}

For the future \yb multi-ion clock on the E3 transition, one challenge is the strong AC-Stark shift of the clock states. Such shifts can be efficiently suppressed by using Hyper-Ramsey spectroscopy techniques \cite{dzubaHyperfineInducedElectricDipole2016}. Nevertheless, the effective excitation of the clock transition in the first and second Ramsey pulses must be guaranteed. Therefore, it is essential to mitigate the differential frequency shift on individual ions induced by the AC Stark shift, which requires a light intensity inhomogeneity of less than 2\% on the Coulomb crystal.

\begin{figure}[h]
	\centering\includegraphics[width=0.45\textwidth]{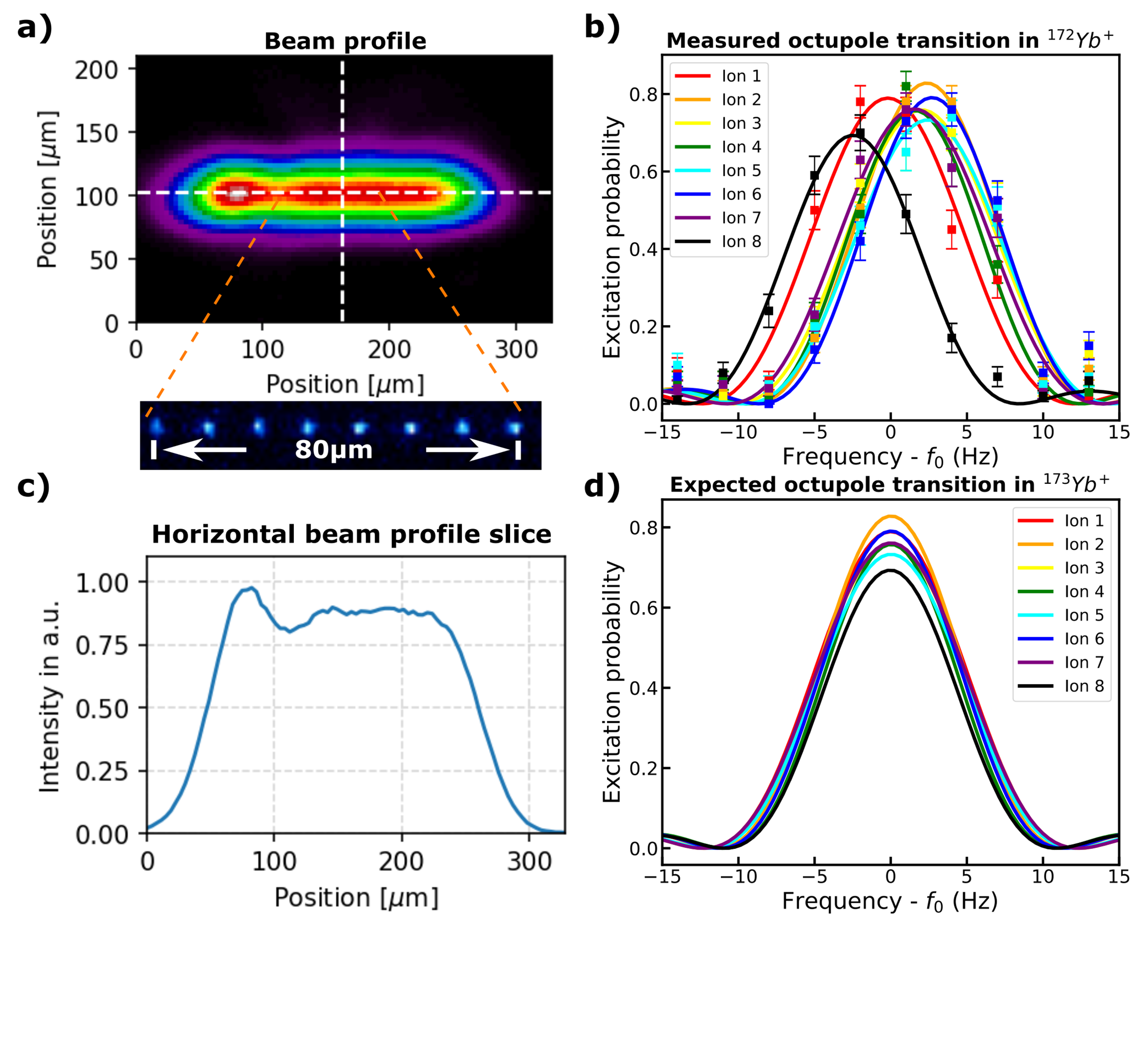}
	\caption{\label{fig3}
a) Intensity profile of the flat-top beam. This beam profile supports simultaneous excitation of multiple \yb ions. b) Frequency scan of the E3 transition on an 8 ion Coulomb crystal (\ybz). c) Horizontal slice of the intensity profile. d) Expected differential frequency of an \ybd Coulomb crystal under the same condition.
        }
\end{figure}
To fulfill the required homogeneity, we utilize a holographic phase plate that generates a flat-top beam profile from a commonly used Gaussian beam. The intensity distribution and a horizontal cut of the generated flat-top beam are illustrated in Fig. \ref{fig3}a and c. This profile could support a Coulomb crystal with a length of 80 µm. With this flat-top beam, we excited 8 \ybz ions on the E3 transition from $^2S_{1/2}$ to $^2F_{7/2}$ state at a Rabi frequency of $2\pi\cdot 6.25$ Hz (Fig. \ref{fig3}b). 

The \ybd is predicted to have nuclear spin quenched, 100 times shorter lifetime of the E3 transition \cite{dzubaHyperfineInducedElectricDipole2016}, and therefore significantly reduced AC-Stark shifts. Utilizing \ybd for the same ion crystal, the differential frequency shift could be significantly reduced as shown in Fig. \ref{fig3}d. With this reduced AC Stark shift, the same beam profile could support multi-ion clocks with 20 ions.

\section{Conclusion and outlook}

We characterized three effects that could degrade the multi-ion operation of \yb ions. The achieved homogeneity of the magnetic field and the RF field supports future multi-ion clock and LLI tests. We also demonstrated coherent excitation of an 8-ion-crystal (\ybz) on the E3 transition with a flat-top beam. With the demonstrated beam profile, an optical clock based on \ybd could support multi-ion operation with 10s of ions, which relaxes the requirement on the local oscillator (ultra-stable laser). The scalable ions traps developed in our group allow for simultaneous clock operation of multiple clock ensembles of a few ten ions each \cite{herschbachLinearPaulTrap2012b, kellerControllingSystematicFrequency2019a}. With the integration of novel technologies such as zero-dead-time clocks \cite{biedermannZeroDeadTimeOperationInterleaved2013a} and cascaded clocks \cite{borregaardEfficientAtomicClocks2013b}, ion-based clocks with significantly enhanced frequency stability or capable of operating in harsh environments can be anticipated.

\begin{acknowledgments}
This project has been funded by the Deutsche Forschungsgemeinschaft (DFG, German Research Foundation) under Germany’s Excellence Strategy – EXC-2123 QuantumFrontiers–390837967 (RU B06) and through Grant No. CRC 1227 (DQ-mat, project B03). This work has been supported by the Max-Planck-RIKEN-PTB-Center for Time, Constants and Fundamental Symmetries. L.S.D. acknowledges support from the Alexander von Humboldt foundation.

\end{acknowledgments}%

%
\end{document}